\documentclass[12pt]{article}
\usepackage{scicite}

\usepackage{times}
\usepackage{amssymb}
\usepackage{graphicx}

\newenvironment{sciabstract}{%
\begin{quote} \bf}
{\end{quote}}

\title{Quantum-enhanced sensing of axion dark matter with a transmon-based single microwave photon counter
}

\author
{C. Braggio$^{1,2,\ast}$, L. Balembois$^{3}$, R. Di Vora$^{4}$, Z. Wang$^{3}$, J. Travesedo$^{3}$, L. Pallegoix$^{3}$,\\ G. Carugno$^{2}$, A. Ortolan$^{4}$, G. Ruoso$^{4}$, 
U.~Gambardella$^{5}$, D.~D'Agostino$^{5}$, \\ P. Bertet$^{3}$, E. Flurin$^{3,\ast}$\\
\\
\normalsize{$^{1}$ Dipartimento di Fisica e Astronomia, Padova, Italy}\\
\normalsize{$^{2}$  INFN, Sezione di Padova, Padova, Italy}\\
\normalsize{$^{3}$ Quantronics group, Universit\'e Paris-Saclay, CEA, CNRS, SPEC, 91191 Gif-sur-Yvette Cedex, France}\\
\normalsize{$^{4}$  Laboratori Nazionali di Legnaro, Legnaro, Padova, Italy}\\
\normalsize{$^{5}$  INFN, Sezione di Napoli, Napoli, Italy}\\
\\
\normalsize{$^\ast$To whom correspondence should be addressed};\\ \normalsize{e-mail:  caterina.braggio@unipd.it, emmanuel.flurin@cea.fr.}
}

\date{}

\begin{document} 
\baselineskip24pt

\maketitle 
\begin{sciabstract}

We report an axion dark matter search with a haloscope equipped with a microwave photon counter. The haloscope is a tunable high quality factor 3-dimensional microwave cavity placed in a magnetic field. The photon counter, operated cyclically, maps an incoming microwave photon onto the state of a superconducting transmon qubit. The measurement protocol continuously monitors the power emitted by the haloscope cavity as well as the dark count background, and enables tuning of the cavity frequency to probe different axion masses. 
With this apparatus we enhance by a factor 20 the search speed that can be reached with quantum-limited linear amplifiers, and set a new standard for probing the existence of axions with resonant detectors.
\end{sciabstract}

\section*{Introduction}

The cosmological model that best complies with the astronomical observations collected over years \cite{Primack:2012aa} invokes the existence of dark matter (DM), having observable gravitational effects but interacting weakly with ordinary matter.  
For decades, experiments with increasing scale and cost have been deployed relying on the hypothetical interaction of DM with the particles of the standard model as devised in specific theoretical models. These tests have set stringent constraints to the concept of the weakly interacting massive particle \cite{Bertone:2018aa} in the mass range 1\,GeV - 10\,TeV \cite{Bertone:2010aa,Liu:2017aa}.
On the ultralight side of the open parameter space, the cold-dark-matter paradigm is pursued with a radically different approach \cite{demille:2017,Sushkov:2023aa}. At this frontier of fundamental physics, diverse small-scale experiments rely on quantum sensing \cite{Degen:2017aa} to improve their sensitivity to new particles and forces in specific mass ranges, with a large discovery potential. 

A prominent example is the cavity axion haloscope, a detector consisting of a 3D microwave resonator permeated by an intense magnetic field and readout by an heterodyne receiver \cite{Sikivie:2021aa}. 
The extremely weak interaction of the axion with electromagnetism is parametrized by $g_{a\gamma\gamma}$ in the Lagrangian $\mathcal{L}=g_{a\gamma\gamma} a \mathbf{E}\cdot \mathbf{B}$, where $a$ is the axion field, and $\mathbf{E}$ and  $\mathbf{B}$ are respectively the electric and magnetic fields. This coupling allows the axion to decay to a photon in a static magnetic field $\mathbf{B}=B_0$, with a conversion rate scaling with $B_0^2$.
The cavity frequency $\nu_c$ sets the particle mass $m_a=h\nu_c$ ($10\,\mu$eV$\simeq 2.5$\,GHz) at which the signal is resonantly enhanced, therefore wide tunability is a main requirement beyond sensitivity.
For axions, best sensitivity results are reported in exclusion plots around a sweet spot for state-of-the-art cavity and magnet technology, in the frequency range around 600\,MHz to 1\,GHz where signal and noise power compare favorably.  Here, SQUID technology and quantum-limited linear amplifiers have enabled probing axion-photon couplings down to $3 \times 10^{-15}$\,GeV$^{-1}$ \cite{Collaboration:2021aa,Yi:2023aa} predicted by quantum chromodynamics (QCD) axion models. However, a large part of the parameter space
 is yet unexplored at higher frequencies as the scan rate decreases with the search frequency as $\nu_c^{-4}$ (see Methods).
Even in the most favorable conditions, assuming haloscopes equipped with the best superconducting magnets delivering fields up to 14 T and state-of-the-art superconducting cavities \cite{Ahn:2022aa} with linewidths $\Delta \nu_c$ matching that of the axion signal $\Delta \nu_a=\nu_c/10^6$ \cite{Turner:1986aa}, the time needed to scan the 1-10\,GHz decade with relevant sensitivity can be estimated around hundreds of years if a detection chain based on the measurement of the field quadratures with a linear amplifier is employed (See Methods). The limiting factor arises from the vacuum state not being an eigenstate of the quadrature operators, which causes the output standard deviation to reach at best an effective noise temperature of half a photon (effective mean occupation number $\bar{n}_{\rm SQL}=1$); this is called the Standard Quantum Limit (SQL). 

Here, we target a significant scan rate reduction in axion DM search by circumventing SQL noise through quantum technologies.
Injection of squeezed states of light has further expanded the probed volume of the gra\-vi\-tational-wave Universe \cite{Collaboration:2019aa,Tse:2019aa}, while in DM search microwave squeezing has allowed for improving the search rate by a factor of 2 \cite{Backes:2021aa,Malnou:2019aa}.
A larger scan rate improvement can in principle be obtained through single microwave photon detectors (SMPDs) \cite{Lamoreaux:2013}, which escape SQL noise since the vacuum is an energy eigenstate.
The scan rate enhancement $\mathcal{R}$ obtained by using a SMPD compared to a quadrature detection at the SQL can be shown to be given by $\mathcal{R} = \eta^2 \Delta \nu_a /\Gamma_\mathrm{dc}$, where $\eta$ is the SMPD quantum efficiency and $\Gamma_\mathrm{dc}$ is the SMPD dark count rate (see Methods). 

A lower bound for the SMPD dark count rate is given by \(\Gamma_{\mathrm{dc}}/(\eta \Delta \nu_c) = \bar{n}_{\mathrm{th}}= 1/\left(e^{h\nu/k_B T}-1\right) = 2.4 \times 10^{-8}\), for an experiment at \(T=20\) mK and frequency \(\nu=7.3\) GHz. However, present-day devices do not reach this figure due to technical issues such as inefficient thermalization of the microwave field at millikelvin temperatures and the presence of out-of-equilibrium quasiparticles in superconductors \cite{Diamond:2022, Connolly:23, balembois2023}. Nonetheless, even with realistically achievable dark count rates \(\Gamma_{\mathrm{dc}} = 10\) s\(^{-1}\) and \(\eta = 0.8\), scan rate enhancements \(\mathcal{R} \sim 500\) can be obtained for state-of-the-art cavities with a loaded quality factor \(Q_L \sim Q_a=\nu_a/\Delta \nu_a =10^6\). Reducing the operational dark count rate $\Gamma_{\mathrm{dc}}$ is therefore essential to achieve a fast scan rate.

The potential of microwave photon counting has been recently demonstrated in the field of dark matter search with quantum non-demolition measurements of cavity photons, using a device based on a transmon qubit coupled to the signal 3D cavity and to another 3D cavity for signal readout \cite{Akash:2021}. 
 In this method, the noise has been reduced to the background photons $\bar{n}_{\rm th}=7.3\times 10^{-4}$ compared to $\bar{n}_{\rm SQL}=1$ at SQL, corresponding to a projected scan speed enhancement by a factor $\sim1300$.
 For axion search however, a large magnetic field must be applied to the axion cavity, which makes it difficult to integrate the transmon. In this work, we take a different approach by spatially separating the axion cavity from the detector, using a transmon-based SMPD which can count photons propagating in a coaxial cable. This makes it possible to perform an axion search in magnetic fields up to 2T, while still obtaining state-of-the-art dark count rates of less than 100\,s$^{-1}$.

\section*{Setup and protocol}

\begin{figure}[]
\centering
\includegraphics[width=0.7\textwidth]{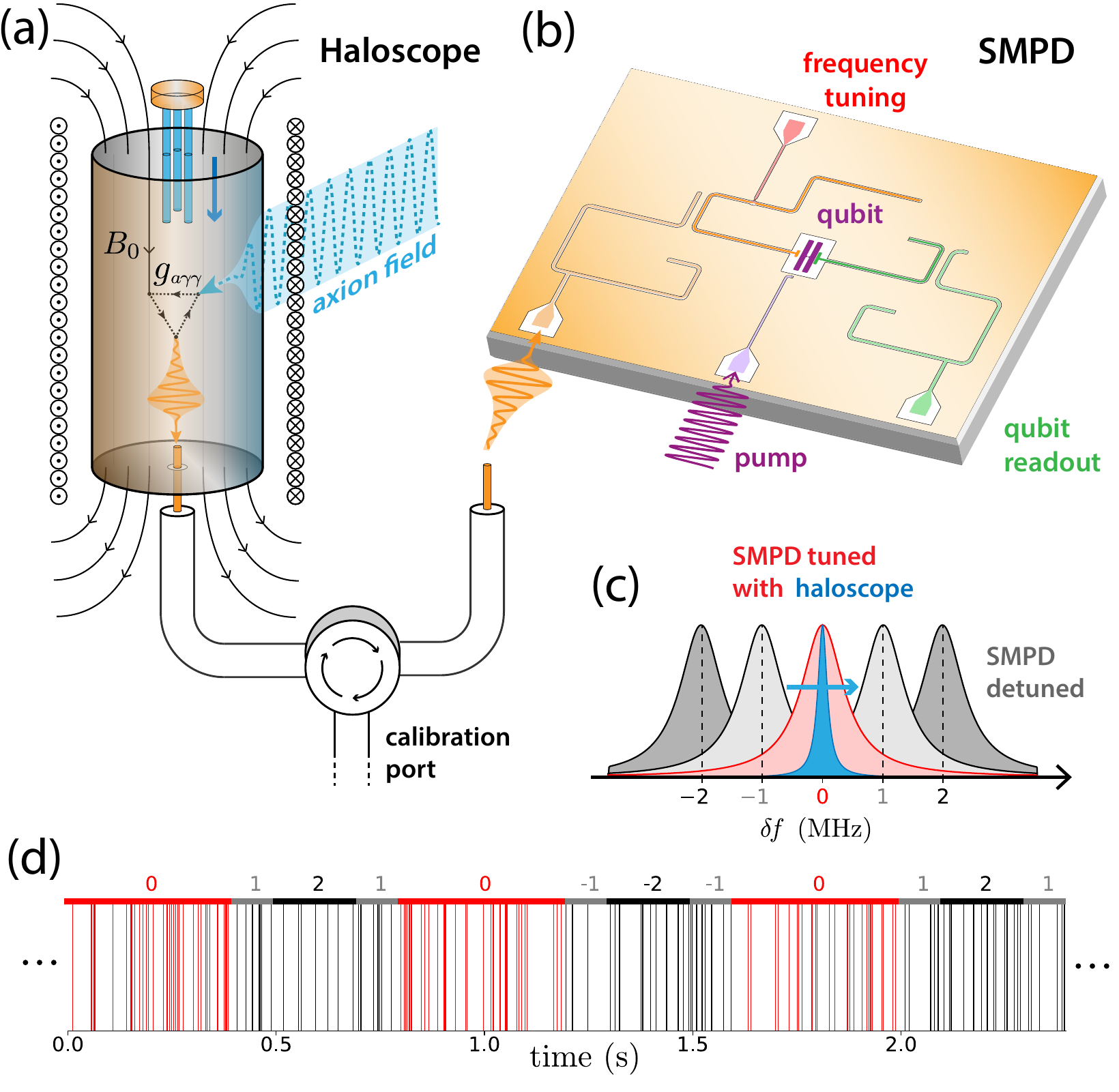}
\caption{Schematic of the axion search setup: (a) The haloscope cavity, located in a 2T magnet, connects to the detector via a fixed antenna port and features cryogenic frequency tuning through three sapphire rods attached to a nano-positioner. (b) The SMPD, a superconducting circuit with $\lambda/2$ coplanar waveguide resonators linked to a transmon qubit, is positioned approximately 50 cm above the magnet and connects via standard coaxial cables. Its frequency is adjustable by threading the flux through a SQUID embedded in the buffer resonator. Upon activating the four-wave mixing process, the qubit cycles through photon detection phases. (c)  The detector center frequency alternates between resonance (red) and off-resonance (grey) settings relative to the haloscope's frequency (blue) in differential mode. (d) Measurement records from the photon counter display clicks over time, with color indicating the detector's frequency setting. 
\label{fig:1}}
\end{figure}

The present haloscope is based on a hybrid surfaced cylindrical NbTi-copper cavity \cite{Alesini:2019}, mounted to the base stage of a dilution refrigerator at 14\,mK. 
At the maximum applied field of 2\,T we measured $Q_0=0.9\times 10^6$ for its axion-sensitive TM$_{010}$ mode. Its frequency can be varied within a few MHz around $7.37$\,GHz  by a system of three 1\,mm-diameter sapphire rods controlled by a cryogenic nanopositioner. The cavity TM$_{010}$ mode is readout by a fixed antenna with coupling coefficient $\beta=3$, thus the loaded quality factor is $Q_L=Q_0/(1+\beta)=2.25\times 10^5$ in the overall frequency range investigated in the present work.
The expected axion signal power is at the $10^{-24}\,$W level (see Methods).  
A circulator routes pulses from input lines towards the cavity for calibration of the cavity parameters, and the photons leaking out of the cavity towards the input of the SMPD (see Fig.~\ref{fig:1}a).\\ The SMPD is a superconducting circuit with a transmon qubit coupled to two resonators: a \lq buffer' resonator whose frequency $\omega_b$ can be tuned to the incoming photon frequency by applying a magnetic flux to an embedded superconducting quantum interference device (SQUID), and a \lq waste' resonator with fixed frequency $\omega_w$.  In this circuit, an itinerant photon entering the buffer resonator is converted with close to unit efficiency into a qubit excitation $\omega_q$ and a waste photon via a four-wave-mixing (4WM) process \cite{Lescanne:2020aa} activated by a pump pulse at frequency $\omega _p$ such that $\omega_p+\omega_b=\omega_q+\omega_w$, corresponding to the energy conservation throughout the 4WM process. The waste resonator quickly damps its converted photon in the environment ensuring the irreversibility of the 4WM process and forcing the qubit to remain in its excited state. The qubit state is then measured using the dispersive readout method \cite{Gambetta:2006}. If it is found in its excited state, then a click of the detector is recorded and the qubit is reset in its ground state. These operations are repeated in cycles of $17\ \mathrm{\mu s}$ on average yielding to measurement records of click arrival time as displayed in Fig.1d. This detector also finds applications in magnetic resonance \cite{Albertinale:2021aa} and in particular it recently enabled single-electron-spin detection \cite{wang2023}.

\begin{figure}[!tbh]
\includegraphics[width=.5\textwidth]{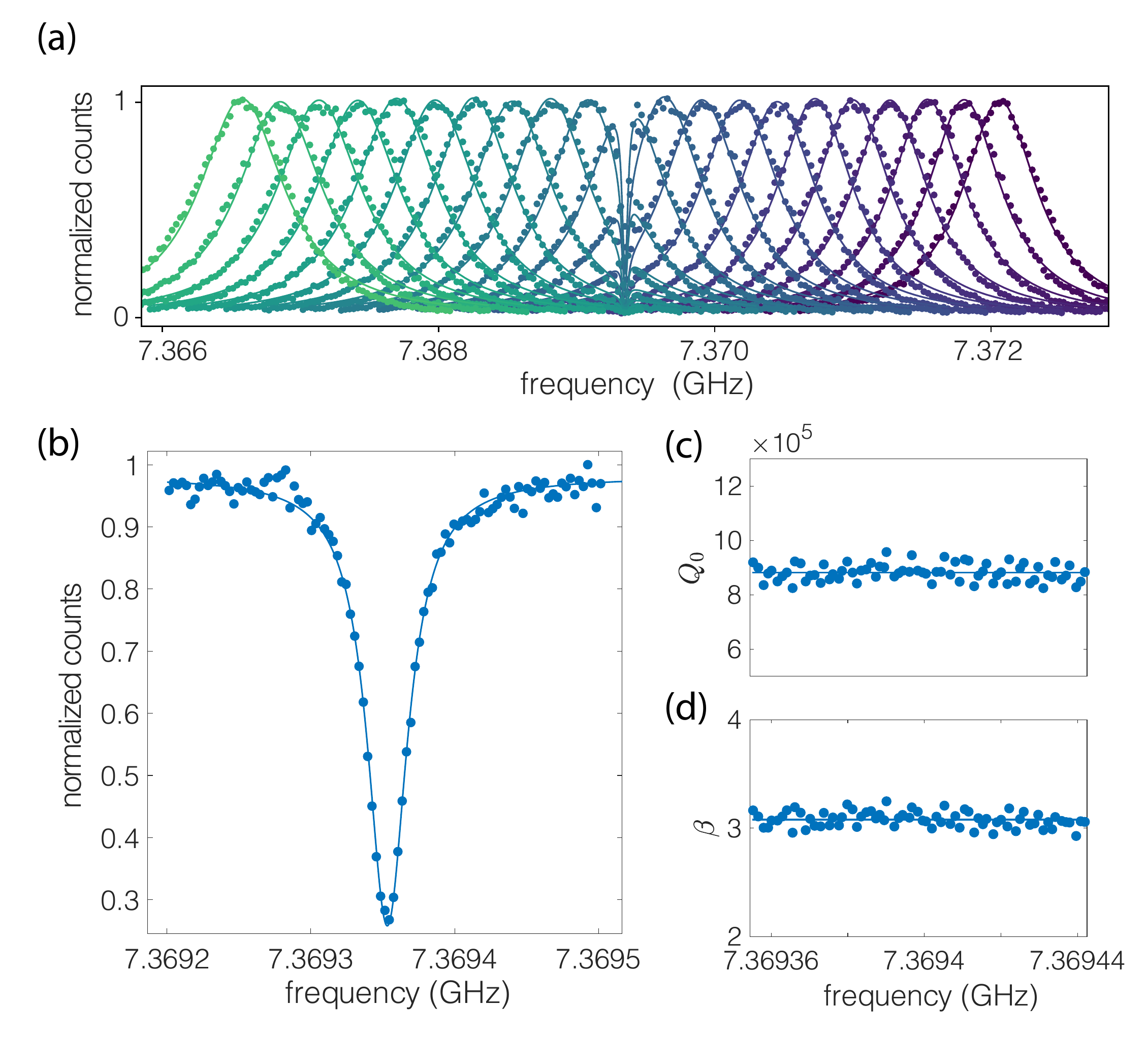}
\centering
\caption{
Microwave cavity spectroscopy by a photon counter. (a) SMPD response to a continuous calibration tone as function of its frequency. While the SMPD bandwidth is $0.7\ \mathrm{MHz}$, its center frequency is tunable over a frequency range of over $100\ \mathrm{MHz}$ by threading magnetic flux in the SQUID loop. Here the SMPD is tuned over a  $5\ \mathrm{MHz}$-span around the haloscope frequency that appears as a dip in the measured counts. (b) The SMPD frequency is swept around haloscope resonance to estimate the cavity parameters $Q_0$ and $\beta$.
(c-d) Measurements of the cavity parameters are repeated cyclically within the quantum protocol upon cavity frequency scanning. For the interval $[7.369355-7.369442]$\,GHz, the quality factor is $Q_0=(8.8135\pm 0.31) \times 10^5$ and the coupling coefficient is $\beta=3.08\pm0.07$, therefore independent of the cavity frequency to within a few percent. The average value of $Q_0$ and $\beta$ for the $N=72$ measurements are represented by the horizontal lines.
}
\label{fig:2}
\end{figure}

Axion dark matter experiments search for a power excess above a background that must be reliably estimated at each cavity frequency. We estimate the background by recording off-resonance clicks when $\omega_b$ is detuned by 1 and 2\,MHz, and the source plus background when $\omega_b=2\pi\nu_c$. To explore different axion masses, the cavity frequency $\nu_c$ is tuned during the quantum sensing protocol, with parameters $\nu_c$, loaded quality factor $Q_L$ and coupling $\beta$ to the transmission line being monitored periodically with the SMPD (see Fig.\,\ref{fig:2}).

The detector efficiency $\eta$ is measured by monitoring the count rate of the detector while applying a microwave tone at its input, whose power is calibrated by measuring the AC-stark shift and photon-induced dephasing of the transmon qubit (see methods). We measure an operational efficency $\eta=0.47 \pm 0.013$ on average including dead times and imperfections of the SMPD. We observe efficiency fluctuations on time scale of minutes of the order of $\pm 10\,\%$ mainly due to slow drifts of the flux threading the detector SQUID loop as well as fluctuations of the relaxation time fluctuation of the transmon qubit \cite{balembois2023}.

\section*{Data analysis and SMPD diagnostics}

\begin{figure}[h!tb]
\centering
\includegraphics[width=.87\textwidth]{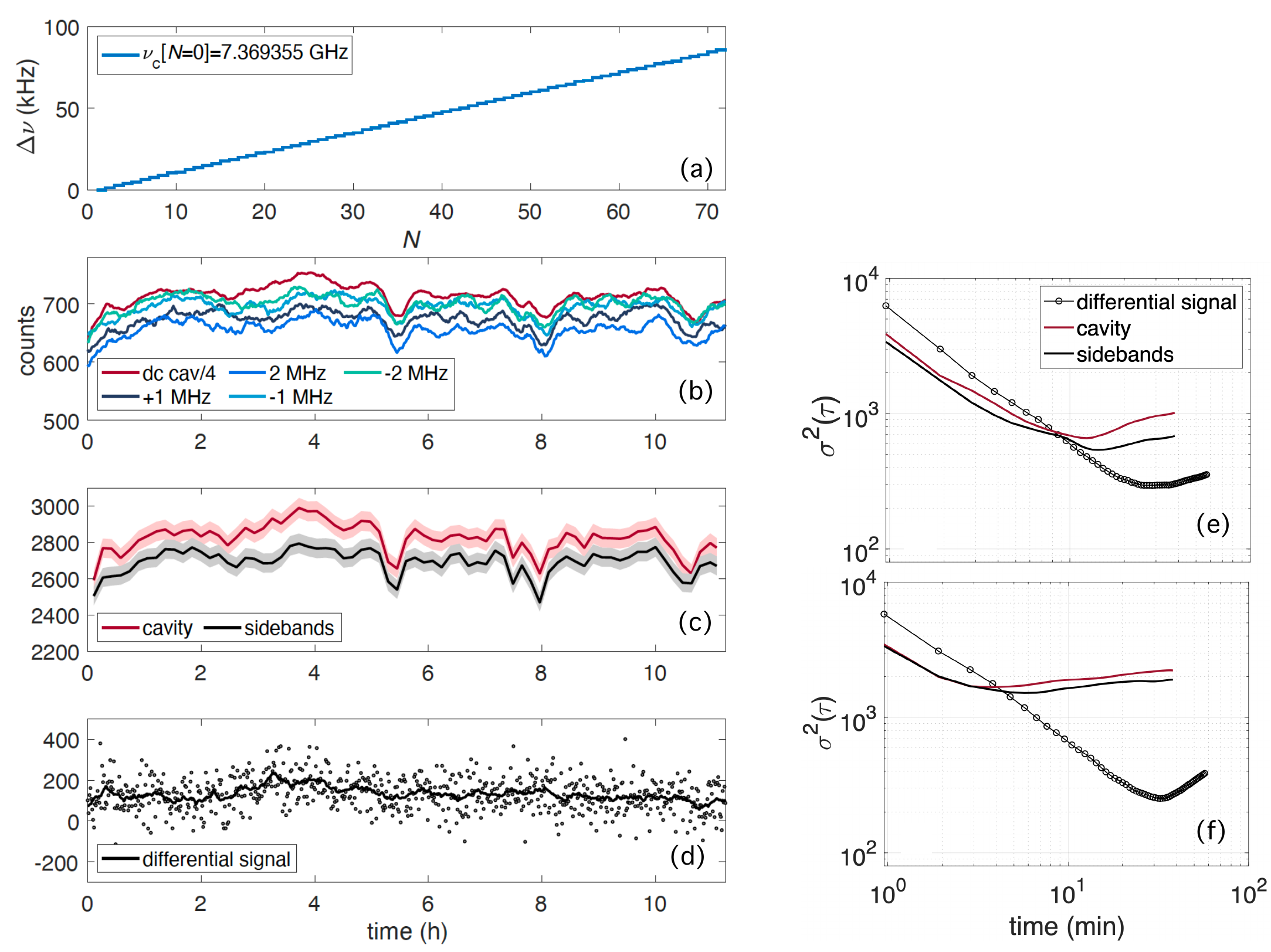}
\caption{
A selection of data taken at about 5\,kHz/h frequency tuning speed, corresponding to the cavity frequency range $[7.369355-7.369442]$\,GHz. (a) The cavity frequency changes linearly with cycle number ($N=72$ in this run). (b) Counts recorded with SMPD buffer $\nu_c\pm 1$\,MHz, $\nu_c\pm 2$\,MHz and at resonance. Each data point represents counts recorded in 7.15\,sec at sidebands frequency. For comparison, the counts recorded at resonance have been divided by a factor 4. (c) Sidebands counts are summed to obtain a single series of data, that is compared to counts recorded at resonance ($\nu_b=\nu_c$). The $\pm \sigma$ belts indicate the statistical uncertainty expected for poissonian counts.
 (d) Excess counts at cavity frequency obtained by the difference between cavity and sideband counts. The thick black curve is obtained by computing the mean over a sliding window of 20\,minutes. 
 (e) Allan variance calculated for data recorded at fixed cavity frequency, and (f) for the data
displayed in (c)-(d). 
}
\label{fig:3}
\end{figure}

The counts acquired as described in the previous section are then grouped into resonance and off-resonance counts. Due to the chosen quantum protocol structure, the acquisition duration at frequency step is 28.6\,s at resonance, and $7.15\,{\rm s}\times 4=28.6$\,s at sidebands frequency. 
Over the long timescales required in cavity haloscope searches, the SMPD dark count rate is non stationary with variations within 10\,\% as shown in Fig.\,\ref{fig:3}\,(b). 
We observe that about 600-700 clicks are recorded in 7.15\,sec on each sideband frequency, while about 2700-2900 on resonance clicks are registered for 28.6\,s-duration intervals. These clicks are originating from photons present at the SMPD input due to an effective temperature of the input line, and only to a minor extent from spurious excitation of the transmon qubit in absence of incoming photons \cite{balembois2023}. 
 The sum of the counts associated with the four sidebands is an estimator of the background $B$, that can be compared with on resonance counts in the same 28.6\,s-duration time window as shown in Fig.\,\ref{fig:3}\,(c), where a correlation between the two measured quantities is evident.  Both the counts registered at cavity frequency (0) and on sidebands (-2,-1,1,2) vary beyond statistical uncertainty expected for poissonian counts, as indicated by the $\pm 1\, \sigma$ belts. 
If $\{t_i\}$ is the set of click arrival times $t_i$, we indicate with  $N_c$ and $N_b$ the number of the cavity and background clicks, respectively.
The difference between the cavity $N_c$ and background counts $N_b$ (see Fig.\,\ref{fig:3}\,(d)) in the overall $0.4$\,MHz probed frequency range is unrelated to any dark matter signal inasmuch as it is independent of the cavity frequency, as detailed in the following. This excess at the cavity frequency is instead ascribed to a slightly higher temperature for the cavity resonator compared to the SMPD temperature. In the data analysis this difference is treated as a bias $k_b$ (see below).

 The SMPD long term stability can be assessed by calculating the Allan variance for $N_c$ and $N_b$ (see Fig.\,\ref{fig:3}). The click number fluctuations is computed as function of the total integration time $\tau$. At early times, the Allan variance decreases as 1/$\tau$ but after only few minutes it increases again, indicating that a random walk contribution to the click rate comes into play.
However, the Allan variance of the difference $N_c-N_b$ follows the $1/\tau$ trend up to a much longer time interval $\tau \sim 30$\,min, suggesting that there are common processes affecting the counter operation. 
Moreover, the plots in Fig.\,\ref{fig:3} show that there is no additional noise in the data recorded between successive step motion intervals compared to those acquired with the cavity unperturbed for several hours.  

\section*{Axion dark matter sensing}

The clicks recorded with SMPD frequency tuned at the cavity frequency and those at sidebands frequency can be used not only to learn about the long-term stability of the SMPD, but also to obtain an upper limit on the axion-photon interaction $g_{a \gamma\gamma}=g_{\gamma}\alpha  f_a^{-1}/\pi$, where $\alpha$ is the fine structure constant and $f_a$ is the scale of the breaking of the Peccei-Quinn symmetry, with  $f_a/10^{12}{\,\rm GeV}=5.691\,\mu{\rm eV}/m_a$ \cite{Chadha:2022}. 

\begin{figure}[h!]
\includegraphics[width=.9\textwidth]{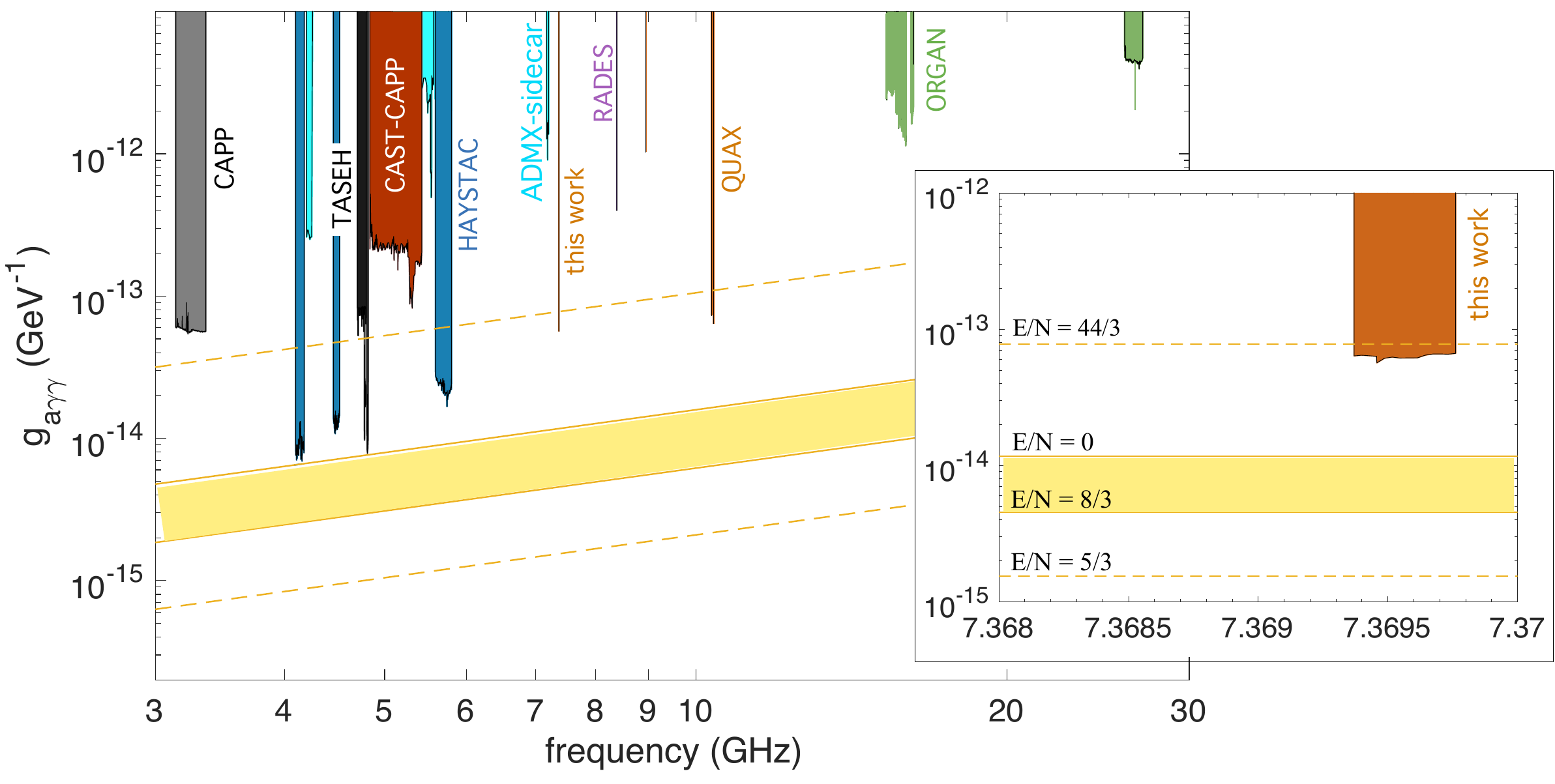}
\caption{(a) Constraints on axion-photon coupling: haloscopes closeup in the range $2-50$\,GHz. Experimental data taken from Ref.\,\cite{cajohare}. The bounds represented by the dotted lines enclose the region of parameter space of phenomenologically preferred axion models \cite{Di-Luzio:2017aa}, extending beyond the commonly assumed QCD axion window (yellow band). E and N are respectively the electromagnetic and QCD anomaly coefficients, setting the axion-photon coupling constant $g_{a\gamma \gamma}=(\alpha/2\pi f_a)(E/N)$ \cite{DILUZIO20201}.
(b) Exclusion limit at 95\% confidence level on the axion mass coupling parameter space. We obtain an upper limit on the axion induced power that translates to an upper limit on g$_{a\gamma\gamma}$ for $m_a \in [30.477, 30.479]\,\mu$eV, corresponding to a cavity frequency window of 0.4\,MHz centred around 7.3696\,GHz. }
\label{fig:7}
\end{figure}

The system has been investigated for 12\,h at fixed cavity frequency, and for an overall range of about 420\,kHz centred at 7.3695\,GHz by tuning the cavity frequency at about $4.5-5$\,kHz/h, as heating introduced at the 15\,mK stage by the currents needed to drive the nanopositioner degraded the SMPD sensitivity for tuning speeds above approximately 12\,kHz/h. 
The probed frequency range corresponds to 14 cavity linewidths, each providing independent values of $N_b$ and $N_c$ useful for the dark matter search. Clearly, the measurement time $\Delta t$ spent probing $[\nu_c,\nu_c+\Delta \nu_c]$, with $\Delta \nu_c=32$\,kHz, largely exceeds $\tau_m$, thus we selected a subset of data to optimize the haloscope speed.
Each $\Delta t$ has been divided into sub-intervals of duration $\Delta t_m=10$\,min and the interval with the maximum SMPD sensitivity, i.e. having the lowest cavity counts $N_c^{\star}$, was selected to devise the plot in Fig.\,\ref{fig:7}.

In each sub-interval, the counts mean equals their variance as expected for Poissonian statististics, therefore to infer the upper limit we can apply the maximum likelihood ratio test to assess the significance of an excess at the cavity frequency \cite{Vianello:2018aa}.
In the recorded data we observe that for the selected minimum cavity counts  $N_c^{\star}$ in the 10\,min duration sub-intervals, the corresponding background counts $N_b^{\star}$ is biased by $k_b=N_c/N_b-1$, which amount to a few percent on average in the collected data set. Using the Wilks' theorem \cite{Vianello:2018aa} for large sample size, as is the case for the present data ($O(10^4)$ counts in each sub-interval),
the significance for $N_c^{\star}$ counts for each cavity linewidth is given by

\begin{eqnarray}
\label{esse}
S&=&\sqrt{2}\left\{N_b^{\star} \log\left[\frac{(k_b+2)N_b^{\star}}{N_b^{\star}+N_c^{\star}}\right]\right.  \nonumber  \\ 
 & + & \left.N_c^{\star} \log\left[\frac{(k_b+2)N_c^{\star}}{(1+k_b)(N_b^{\star}+N_c^{\star})}\right] \right\}^{1/2} ,
\end{eqnarray}
where we have taken the same detection detection efficiency for cavity and background counts.

The required significance to claim a detection is set at 5$\sigma$, corresponding to a false alarm probability of $\sim10^{-7}$ for a Gaussian distribution.
In the four acquisition runs, representing 14 cavity linewidths, no significant excess was found assuming a conservative bias value $k_b=0.05$. Therefore we
set the upper limits  $N_{95}^{\star}$  on source counts at 95\,\% C. L. (i.e. 2$\sigma$ significance),
by interpreting
 excess counts $N_{95}^{\star}-N_b^{\star}$ as signal power $P_{95}$ due to axion to photon conversion:
\begin{equation}
\label{ }
P_{95}=\eta h\nu_c \frac{\,\, N_{95}^{\star}-N_b^{\star}}{\Delta t_m}.
\end{equation}
The upper limit on $g_{a \gamma \gamma}$ obtained from the calculated axion signal power (see Methods) and eq.\,\ref{esse} is reported in Fig.\,\ref{fig:7} for the probed frequency range. 

\section*{Conclusions and outlook}

At sensitivity given by the potential interaction of axions with photons, we devised an axion search protocol whereby clicks are recorded at both the cavity and the sidebands frequency while the cavity frequency is changed much more slowly than the execution time of the protocol. 
Allan variance plots indicate that for the present device the most convenient haloscope integration time is of about 10-15\,min. As the inhomogeneity of the Poisson process sets in for temporal intervals much larger than the chosen integration time of 10\,minutes, we are able to give an upper limit $g_{a \gamma \gamma}<7\times 10^{-14}$\,GeV$^{-1}$ in the mass range $(30.477-30.479)\,\mu$eV.\\
The exclusion plot we obtained corresponds to a search speed of 4.3\,MHz/day. As $\Delta f/\Delta t_m$ is proportional to the $B^4\,g_{a \gamma \gamma}^4\,V_{eff}^2$, with $V_{eff}=C_{010}V$ the effective cavity volume, we can estimate a scan rate exceeding a hundred MHz/year at the sensitivity required to probe the full QCD axion band for an experiment equipped with a commercially available magnetic field of 12\,T in place of the 2\,T magnet used in these tests, and a state-of-the-art 3D resonator with 10 times as much the effective volume. 
Compared to a haloscope based on a SQL linear amplifier, photon counting allows for running the search at a speed larger by a factor of $\mathcal{R}=\eta^2 \Delta \nu_a /\Gamma_{dc} \sim 20$, with the signal linewidth $\Delta \nu_a=7.3\ \mathrm{kHz}$, the dark count rate $\Gamma_{dc}=85\ \mathrm{s^{-1}}$ and the efficiency $\eta = 0.46$. 

The gain in scan speed we have demonstrated is currently limited by the dark count rate, which can be further suppressed by improving the thermalization of the lines or by narrowing the bandwidth of the input resonator. 
In addition, we have devised a detection method applicable to broad frequency ranges, as the SMPD input resonator frequency can very rapidly be adjusted to the cavity frequency while the latter is tuned across a few hundred MHz.  Note that a Josephson mixer \cite{RochFlu2012} allows to virtually expand the bandwidth of the present SMPD to include any cavity haloscope in the interesting, yet unprobed high frequency range where heavier axions can be detected. 
Our results demonstrate the potential of microwave photon counting in axion DM searches above 5\,GHz frequency, accelerating by orders of magnitude the search and significantly simplifying the data acquisition and analysis without losing the required robustness thanks to the employed metrological methods. 

\bibliography{scifile}

\bibliographystyle{Science}

\section*{Acknowledgments}

We are grateful to E. Berto (University of Padova and INFN) who substantially contributed to the mechanical realization of the cavity and of its tuning system. We thank the sputtering laboratory of the INFN Legnaro Laboratory (C. Pira) for the preparation of the NbTi films on the copper resonators. The contribution of M. Tessaro and F. Calaon (INFN Padova) to the experiment electronics and cryogenics respectively is gratefully acknowledged. We acknowledge technical support from P.~S\'enat, D. Duet, P.-F.~Orfila and S.~Delprat, from CEA/SPEC and are grateful for fruitful discussions within the Quantronics group. We acknowledge IARPA and Lincoln Labs for providing the Josephson Traveling-Wave Parametric Amplifier.

This research is supported by the U.S. Department of Energy, Office of Science, National Quantum Information Science Research Centers, Superconducting Quantum Materials and Systems Center (SQMS) under the Contract No. DE- AC02- 07CH11359, by the MUR Departments of Excellence grant 2023-2027 \lq\lq Quantum Frontiers'', and by INFN through the QUAX experiment. This project has received funding from the European Union Horizon 2020 research and innovation program under ERC-2021-STG grant agreement no. 101042315 (INGENIOUS)


\section*{Additional information}

{\bf Competing interests} The authors declare no competing interests. \\

\noindent
{\bf Data and code availability} Raw data, analysis codes are available from the corresponding author on reasonable request.  \\

\renewcommand{\thetable}{Supplementary Table}
\renewcommand{\figurename}{Supplementary Figure}
\setcounter{figure}{0}

\newpage
\section*{Supplementary materials}

\subsection{SMPD Circuit fabrication}
First, a 2-inch sapphire substrate is cleaned by dipping it in a 2:1 mixture of $\mathrm{H_2SO_4}$ and $\mathrm{H_2O_2}$ during 20 min. The substrate is then loaded in a sputtering machine where a 60-nm thick tantalum thin film is deposited at $600^\circ$C to favor  growth  in the $\alpha$-phase.  The wafer is then diced into rectangular chips of $10\cdot11  \mathrm{mm^2}$.  The patterning of the main elements of the circuit is achieved by wet etching tantalum with Transene 111  through an optically patterned   AZ1518 resist mask.  The chip is then cleaned by immersing it in successive baths: IPA, acetone and 2:1 mixture of $\mathrm{H_2SO_4}$ and $\mathrm{H_2O_2}$.  Aluminum junctions are made using the Dolan bridge technique. The mask consists of a double layer PMMA (110 nm) - MAA (1100 nm) resist and a discharging   layer  of 7 nm of aluminum, it is then patterned using electron beam lithography (at 30 kV)  .  After the exposure, the discharging layer is first removed by immersing the chip in a $\mathrm{KOH}$ solution ($\mathrm{10 g\cdot L^{-1}}$). The resist is then developed in a  1:3 MIBK/IPA mixture.  The chip is then loaded in an electron beam evaporator to make the junctions. The two aluminum layers are evaporated with $28^\circ$ and $-28^\circ$ angles creating overlaps over well-defined area, controlled by the mask geometry. The first aluminum layer is 35 nm thick and the second 65 nm thick. Between the two deposition steps, the aluminum is oxidized during 5 min by injecting an argon-oxygen gas mixture into the deposition  chamber at a pressure of 10 mbar. Finally, the resist mask is lifted-off  by immersing the chip in an acetone bath.  Then, the junctions are then recontacted to the tantalum circuit with  aluminum patches.  After coating the chip with an optical resist (Microposit S1805), windows overlapping the junctions and the ground plane areas  to be recontacted are opened using optical lithography.  The chip is then loaded in  an electron-beam evaporator. We first perform an ion milling based on an argon ion beam accelerated at 500V. Finally,  a  100 nm  aluminium layer is then deposited on the chip covering the areas  etched by the ion milling step. The resist is then removed.

\subsection{SMPD circuit parameters}

In Table \ref{Tab:2} the measured SMPD parameters are reported.
\begin{center}
\begin{table}[h!]
    \begin{tabular}{ l c r   }
    \hline
     \hline
 Qubit & & \\ 
 \hline
    $\omega_q/2\pi$ & &$6.222\ \mathrm{GHz}$\\ \hline
    $T_1$ & & $17-20\ \mathrm{\mu s}$\\ \hline
    $T_{2}^*$ & & $  28\ \mathrm{\mu s}$\\ \hline
    $\chi_{qq}/2\pi$& & $240\ \mathrm{MHz}$\\ \hline
    $\chi_{qb}/2\pi$&& $3.4\ \mathrm{MHz}$\\ \hline
    $\chi_{qw}/2\pi$&& $15\ \mathrm{MHz}$\\ 
    \hline
     \hline
Waste mode &&      \\ \hline
    $\omega_w/2\pi$&& $7.9925\ \mathrm{GHz}$\\ \hline
    $\kappa_\mathrm{ext}/2\pi$&& $ 1.0
\ \mathrm{MHz}$\\ \hline
    $\kappa_\mathrm{int}/2\pi$&& $ < 100\ \mathrm{kHz}$  \\ \hline
         \hline
Buffer mode &&          \\ \hline
    $\omega_b/2\pi$& &$7.3693\ \mathrm{GHz}$\\ \hline
    $\kappa_\mathrm{ext}/2\pi$ && $0.48\ \mathrm{MHz}$\\ \hline
    $\kappa_\mathrm{int}/2\pi$ && $40\ \mathrm{kHz}$\\ \hline   
    \end{tabular}
    \caption{Measured SMPD parameters.}
    \label{Tab:2}
    \end{table}   
\end{center}

\noindent
\subsection{\bf Quantum Sensing Protocol}

The quantum sensing sequence consists of nested cycles as follows (see also Fig.\,\ref{fig:M1}):

(i) The detection cycle begins with a $10.5\ \mathrm{\mu s}$ pump pulse applied to the qubit port to activate the 4WM process, followed by a $0.8\ \mathrm{\mu s}$ readout pulse directed to the waste port. The qubit is then reset utilizing the real-time feedback capabilities of the system controller
resulting in a latency of $0.7\ \mathrm{\mu s}$ that includes electrical delay, signal processing, and FPGA latency. If the qubit is in its ground state, an additional $0.3\ \mathrm{\mu s}$ waiting time is imposed before the cycle restarts. If the qubit is in its excited state, a $0.2\ \mathrm{\mu s}$ qubit $\pi$-pulse is applied, followed by a $0.8\ \mathrm{\mu s}$ readout pulse. This reset procedure continues until successful. The detection cycle time varies, with an average duration of $12.4\ \mathrm{\mu s}$.

(ii) To change the SMPD frequency, a slow flux ramp is applied for 0.72 ms, followed by a 0.1 ms pause to mitigate potential heating. Subsequently, dark counts are logged over 8001 detection cycles (signal OFF), yielding an average time of 99.075 ms. The detector's efficiency (signal ON) is assessed over 801 detection cycles by evaluating the SMPD count rate with a calibrated microwave pulse in use. This efficiency assessment averages 9.91 ms per cycle.

(iii) Sequence (ii) is repeated for buffer frequencies $\omega_b$ matching the cavity frequency, identified as 0, $\omega_c/2\pi\pm 1$\,MHz (labeled as 1,-1), and $\omega_c/2\pi\pm2$\,MHz (labeled as 2,-2). This arrangement facilitates the recording of dark counts for equal durations when at and away from cavity resonance (0\,0\,1\,2\,2\,1\,0\,0\,0\,0\,-1\,-2\, -2\,-1\,0\,0), thus halving the haloscope detector's duty cycle to 50\%. This sequence design ensures all flux ramps share identical amplitude to circumvent systematics related to spurious heating.

(iv) A minimal nano-positioner voltage pulse is applied, followed by a 5 s waiting period. Then, sequence (iii) is executed $N_r=36$ times. This cycle (iv) spans an average of 78 s, covering sequence loading, data collection, and data saving phases.

(v) Cycle (iv) is reiterated 10 times, interspersed with a calibration sequence lasting 78 s, initiated by a subtle nano-positioner voltage pulse to evenly distribute nano-positioner voltage pulses over time and prevent slow thermal fluctuations. The calibration process is twofold. Initially, the haloscope frequency is determined by gauging the photon count reflected from the haloscope cavity across various illumination frequencies using the SMPD. A Lorenzian fit to the haloscope power absorption curve provides an accurate frequency estimate $\omega_c/2\pi$ of the haloscope, down to sub-kHz precision. Next, the SMPD's central frequency and bandwidth are gauged across different bias voltage levels close to the operational point to adjust for low-frequency magnetic flux drifts. Subsequently, the SMPD's central frequency is aligned with the newly measured haloscope frequency. Additionally, a list of bias voltage values is compiled for adjusting the buffer frequency to $\omega_c$ (0), $\omega_c/2\pi\pm 1$\,MHz (1,-1), and $\omega_c/2\pi\pm2$\,MHz (2,-2).

(vi) Cycle (v) is executed continuously, ensuring a consistent operation flow. The active measurement time dedicated to haloscope observation within cycle (v) amounts to 99.075 ms*36*8*10= 285 s, equivalent to 4.76 minutes. This duration is intentionally mirrored for the SMPD background measurement to maintain the differential mode operation of the SMPD-based axion search. Consequently, the total span of cycle (v) averages to 920 seconds or approximately 15.3 minutes. Within this timeframe, approximately 350 seconds are identified as dead times per cycle, accounting for about 38\% of the total cycle duration. This segment includes critical operations like efficiency verifications and frequency adjustments alongside other less crucial intervals for tasks such as extended waiting periods, data processing, and saving. Opportunities to refine and reduce these dead times will be explored in subsequent experimental setups, aiming for enhanced efficiency and throughput.

\noindent
\subsection{Efficiency Calibration}
The calibration of the SMPD is conducted by measuring the dephasing and AC-Stark shift of the qubit, which is induced by the illumination of the input resonator of the SMPD with a weak coherent tone. This calibration is carried out without any pump tone, and thus the SMPD is treated merely as an elementary component of a circuit quantum electrodynamics system. In this setup, the qubit is dispersively coupled to the input resonator, following the Hamiltonian in the rotating frame of the coherent drive:

\begin{equation}
\hat{H}/\hbar= \Delta\hat{a}^\dagger\hat{a}+ \frac{\omega_\mathrm{q}}{2}\hat{\sigma}_z-\frac{\chi}{2} \hat{a}^\dagger\hat{a}\hat{\sigma}_z
+\epsilon(\hat{a}^\dagger+\hat{a})
\end{equation}
where $\omega_q$ represents the qubit frequency, $\chi$ the qubit dispersive shift, $\Delta$ the detuning between the coherent drive and the resonator frequency, and $\epsilon$ the coherent drive rate of the resonator.

The qubit, influenced by the coherent drive, experiences a frequency shift corresponding to the average number of photons in the cavity, while photon number fluctuations lead to a qubit dephasing also dependent on the photon number.

Following the method detailed in \cite{Gambetta:2006}, the frequency shift $\delta\omega$ and dephasing $\delta\gamma$ are directly linked to the coherent complex amplitudes $\alpha_{g}$ and $\alpha_e$ of the photon field inside the resonator induced by the coherent drive, for the qubit in its ground and excited states respectively, as given by the expression:

\begin{equation}\label{eq: photon induced dephasing}
    {\delta\omega+i\delta\gamma=-\chi \alpha_g\bar{\alpha_e}=\frac{-4\chi |\epsilon|^2}{(\kappa+i\chi)^2+4\Delta^2}}
\end{equation}

where $\kappa$ is the dissipation rate of the resonator. The coherent complex amplitudes are defined as:

\begin{equation}\label{eq: photon induced dephasing2}
    \alpha_{g/e}=\frac{\epsilon}{\kappa/2+i(\Delta\mp\chi/2)}
\end{equation}

The qubit frequency shift and dephasing are experimentally measured by performing a Ramsey experiment on the qubit, varying the frequency of the coherent drive applied to the cavity. The measured quantities, which include qubit decay rates and frequency shift, are plotted as a function of the frequency detuning of the coherent drive. The dispersive shift $\chi$ and the cavity decay rate $\kappa$ are extracted from the overall shape of the curve, with $\epsilon$ serving as the frequency scaling parameter.

The power of the coherent drive at the input of the cavity is calculated using input-output relations:

\begin{equation}
P_\mathrm{in} = \hbar \omega_\mathrm{0} \kappa \frac{|\epsilon|^2}{(\kappa-\kappa_\mathrm{l})^2}
    \label{eq:input_power}
\end{equation}
where $\kappa_\mathrm{l}$ is the rate of the resonator loss channel, measured independently through reflectometry using a vector network analyzer. The power input, expressed in units of photon flux, is $P_\mathrm{in}/\hbar\omega$. The error bars are computed from statistical uncertainty associated from the data shown in Fig.\ref{fig:S2}.

As illustrated in Fig.\ref{fig:S2}, this calibration process determines that the photon flux at the input of the resonator for the calibration coherent tone is $P_\mathrm{in}/\hbar\omega=20 050\pm 340 \ \mathrm{photon.s^{-1}}$, or in power units, $P_\mathrm{in} = 0.979\pm0.016 \times 10^{-19} \ \mathrm{W}$.  

With the input tone calibrated, the operational efficiency of the SMPD, based on the 4-wave-mixing process and including duty cycles and dead times, is assessed by measuring the click rate of the device over a minute while continuously illuminated with the calibrated coherent tone. This process reveals an excess click rate over the background dark count of $9233\pm 30\ \mathrm{click.s^{-1}}$, leading to an efficiency of $\eta=0.460 \pm 0.009$. Note that due to fluctuations in the relaxation time of the qubit typical of transmon circuits, a slow drift of the efficiency is expected on hour timescales, this is why the count rate is continuously monitored during the cycle (ii) of the haloscope measurement.

\subsection{Basic Characterization of the SMPD}
The readout of the SMPD utilizes the standard dispersive readout method. A $0.8\ \mathrm{\mu s}$ coherent pulse is directed onto the waste resonator at its resonant frequency when the qubit is in its excited state ($\omega_\mathrm{w}-\chi_\mathrm{qw}$). The reflected signal, encoding the qubit state's complex amplitude, is amplified by a Josephson Parametric Travelling Wave Amplifier (JTWPA) at base temperature, followed by a High Electron Mobility Transistor (HEMT) amplifier at 4K and a low noise amplifier at room temperature. This signal is then demodulated and integrated into a voltage reading, which is compared against a threshold to determine a click event in the SMPD. Fig.\ref{fig:S1} displays the integrated voltage as a function of the qubit state. From this data, we determine the readout fidelity of the qubit when it is in its excited state as $p(1|e)=0.93$, and the thermal equilibrium population of the qubit in its resting state as $p_{th}=2\times10^{-4}$. Additionally, the readout fidelity for the ground state is estimated through consecutive measurements, revealing that the probability of incorrectly identifying the qubit's ground state is less than $5\times10^{-5}$.

The four-wave mixing process is calibrated by measuring the probability of the qubit's excited state while simultaneously illuminating the detector with a pump tone at the matching frequency for four-wave mixing and a coherent tone at the input resonator frequency. As shown in Fig.\ref{fig:S2}a-b, by adjusting the frequencies of these two tones, we identify the optimal conditions where the excited state probability is maximized. We further confirm the accuracy of the mixing process by observing the disappearance of the excited state population when the coherent drive at the input resonator is deactivated. The input resonator's frequency is made adjustable by incorporating a Superconducting Quantum Interference Device (SQUID) at its voltage node. The resonator frequency is modulated by applying a magnetic flux through the SQUID. This adjustment ensures that the resonator's frequency aligns with the haloscope frequency for effective four-wave mixing. The presence of a dark line in Fig.\ref{fig:S2}a, indicative of the absorption of photons resonant with the haloscope, validates that the SMPD is correctly tuned to the haloscope.

\subsection{Frequency Calibration Routine}
This section details the calibration routine executed in each cycle (v), performed every 920 seconds and typically lasting 78 seconds.

\noindent
\subsubsection{SMPD Frequency Routine}
In the frequency calibration routine, we assess the excited state population of the qubit as it relates to the frequency of the coherent drive on the input resonator, spanning $3\ \mathrm{MHz}$.\\
A $2\ \mathrm{mV}$ span voltage bias is applied to control the magnetic flux through the SQUID, thereby altering the input resonator frequency by approximately $\pm 3\ \mathrm{MHz}$. The results, as depicted in Fig.\ref{fig:S3}, are presented in a colorplot, detailing the relative detuning in comparison to the expected frequency of the input resonator. In Fig.\ref{fig:S3}c, the data are shown based on the absolute driving frequency of the input resonator. Key detuned SMPD bias points, such as $\pm1\ \mathrm{MHz}$ and $\pm2\ \mathrm{MHz}$ (targeted for background measurements), are highlighted by vertical lines. The dataset demonstrates a consistent response of the SMPD across a broad frequency range. Through meticulous analysis, we adjust the pump and flux bias for each scanning frequency, countering slow magnetic drifts in the SQUID bias, which is recalibrated every 15 minutes.

\noindent
\subsubsection{Haloscope Frequency Routine}
As established earlier, the SMPD's capability to scan frequencies beyond its linewidth, while maintaining steady efficiency, enables reflection spectroscopy of the haloscope, independent of the SMPD's linewidth. In this process, shown in Fig.\ref{fig:S3}\,b, the SMPD's click rate is recorded while the coherent drive frequency is scanned across a $0.3\,\mathrm{MHz}$ range centered on the haloscope's anticipated resonance frequency. This method ensures a uniform SMPD response as both the SQUID bias and pump frequency are concurrently adjusted. The resultant detailed absorption spectrum of the haloscope is fitted with a Lorentzian curve, from which we extract the haloscope's frequency and linewidth with remarkable sub-kHz precision.

\subsubsection{Resolving the Beta Factor Ambiguity of the Haloscope}
The SMPD's reflection spectrum exclusively reflects the magnitude of the reflection coefficient, leaving an inherent ambiguity: identical spectra can result from distinct beta factors, signifying either an overcoupled ($\beta>1$) or undercoupled ($\beta<1$) state. To resolve this uncertainty, one could analyze the pulsed response of the cavity over time and across frequencies. However, the SMPD's time resolution, limited by its $10\,\mathrm{\mu s}$ detection window, necessitates a reduction in the detection window duration to $1\,\mathrm{\mu s}$ to improve resolution, albeit at the cost of reduced duty cycle ($28\%$). Fig.\ref{fig:S4} showcases the SMPD's click rate in response to a $80\,\mathrm{\mu s}$ pulse applied to the haloscope, with the excitation pulse frequency spanning a $200\ \mathrm{kHz}$ range centered around the haloscope's frequency, and the SMPD frequency adjusted accordingly. This measurement conclusively indicates that the haloscope's temporal response is consistent only with an overcoupled configuration ($\beta=3.1>1$).

\noindent
\subsection{Microwave 3D resonator}
 The 3D resonator where axions might convert to photons is a clamshell cavity of cylindrical body 128\,mm-long and with 31.64\,mm diameter, closed with 10\,mm-long
conical end caps to reduce current dissipation at copper interfaces. The two halves are machined from oxygen-free high thermal conductivity copper, treated with electrochemical polishing before deposition of a superconductig NbTi film by magnetron sputtering. 
As a type-II superconductor, below its critical temperature NbTi is not in the Meissner state but rather in the vortex state, with partial penetration of the magnetic flux in the material. The dissipation mechanism is vortex motion, thus to minimize the overall cavity surface resistance $R_s$ only the cylindrical body, where the currents of the mode TM$_{010}$ are parallel to the applied field, is covered by the NbTi film.  
The internal quality factor $Q_0$, inversely proportional to $R_s$, deteriorates with increasing static B field amplitude as shown in Fig.\,\ref{fig:M2}, where measurements of $Q_0$ made with a twin cavity are reported. This cavity differs from the one used for the present axion search in its diameter, which was slightly larger (diameter $\phi=33.2$\,mm, with measured TM$_{010}$ mode frequency $\nu_c=6.991$\,GHz at room temperature).
In the measurements shown the cavity was mounted in a flow cryostat, in which the temperature of the He flow, kept at a pressure of about 600\,mbar, is controlled with a thermostat down to 3.5\,K. The cavity is inserted in the bore of a solenoid magnet capable of delivering fields exceeding 10\,T. 
Values reported are obtained by doing temperature measurements up to the NbTi critical temperature at zero-field and then by cooling down the cavity to the minimum temperature before increasing the magnetic field. 

\subsection{Axion signal power}
QCD axion models models can be probed with cavity haloscope detectors \cite{Yannis:22,Sikivie:2021aa} in a range from a few hundred MHz up to about 50\,GHz, corresponding to $m_a=h\nu_a=200\,\mu$eV axion mass. 
 In addition to assuming the existence of axions as exclusive constituent of the Galactic dark matter halo \cite{Turner:1986aa}, detectors rely on their resonant conversion into excitations of a high-quality factor electromagnetic cavity mode, whose electric field is parallel to an applied intense magnetic field. 
As is the case for the experiment described in this work, the fundamental mode TM$_{010}$ of an empty copper cylinder resonator is typically used, and the signal power $P_{a}$ is proportional to $(VC_{010}Q_L)$, with $C_{010}=|\int_V dV \mathbf{E}_{010}(\mathbf{x},t)\cdot \mathbf{ B}(\mathbf{ x})|^2/(\mathbf{B}^2 V \int_V dV \mathbf{E}_{010}^2)$, $\mathbf{B}$ external field, $\mathbf{E}_{010}$ is the microwave cavity electric field, $V$ and $Q_L=Q_c/(1+\beta)$ respectively volume and loaded quality factor of the cavity. $\beta$ is the coupling strength of a coaxial  antenna to the cavity mode.
In natural units the axion power extracted with the antenna at resonance is given by \cite{brubaker:2017}:
\begin{equation}
\label{axpow}
P_{a}=g_{\gamma}^{2} \frac{\alpha^2}{\pi^2}\frac{\rho_a}{\Lambda^4} 
\frac{\beta}{1+\beta}\omega_c B_0^2VC_{010}\frac{Q_aQ_L}{Q_a+Q_L}
\end{equation}
\noindent
where $\alpha$ is the fine-structure constant, $\rho_a\simeq 0.45$\,GeV/cm$^3$ is the dark matter density in the galactic halo, $\Lambda=78\,$MeV a parameter linking the axion mass to hadronic physics, and $g_{\gamma}$ is the dimensionless axion-photon coupling. The coupling that appears in the axion-photon Lagrangian is $g_{a\gamma\gamma} =(g_{\gamma}\alpha/\pi \Lambda^2)m_a$. 
A useful benchmark for experiments is the QCD axion band, delimited by the KSVZ (Kim-Shifman-Vainshtein-Zakharov)\cite{Kim:1979aa,SHIFMAN1980493} and DFSZ (Dine-Fischler-Srednicki-Zhitnitsky)\cite{DINE1981199,Zhitnitsky:1980} families of models, with couplings $g_{\gamma}=-0.97$ and 0.36 respectively. 
A practical expression for the signal power is:
\begin{equation}\label{axpow2}
\begin{array}{r@{}l}
P_{a\gamma\gamma} &{}= 0.72 {\rm yW} \left(\frac{g_{\gamma}}{0.97}\right)^2\frac{\rho_a}{0.45\,{\rm \,GeV/cm}^3}\left(\frac{B}{2\,{\rm T}}\right)^2\cdot\\
&{}\,\,\, \cdot \frac{V}{0.11\,l} \cdot \frac{\nu_c}{7.37\,{\rm GHz}} \cdot \frac{Q_L}{225000} \cdot \frac{C_{010}}{0.64},
\end{array}
\end{equation}
in which the parameters of the present experiment are made explicit. 
The smallness of the signal power, which at DFSZ models is at the yoctowatt ($10^{-24}$\,W) level even when state-of-the-art equipment is employed (see table \ref{t1}), is thus the key methodological challenge for haloscope detectors.
\begin{table}[ht]
\centering
\begin{tabular}[t]{l|c|c c c}
\hline
& B [T] & $P_{a}^{\rm KSVZ}$\,[yW(ph/s)]  & \, & $P_{a}^{\rm DFSZ}$\,[yW(ph/s)]  \\
\hline
$\nu_c=7.37$\,GHz & 2 & 0.84(0.17) & & 0.11(0.026)\,\\
 & 12 & 30.4(6.2) & & 6.3(0.86)\\
\hline
\hline
$\nu_c=10$\,GHz & 12 & 22.39(3.38) & & 3.11(0.47)\\
\hline
\end{tabular}
\caption{Signal power for benchmark QCD axion models in yoctowatt (yW$=10^{-24}$\,W) and photon rate calculated with eq.\ref{axpow}. 
The right cylinder hybrid surfaced cavity employed in the present experimental apparatus has a fundamental frequency of resonance $\nu_c=\,7.37$\,GHz ($30.48\,\mu$\,eV axion mass),  volume $V=0.1\,$l, form factor $C_{010}=0.64$, and loaded quality factor $Q_L=2.25\times10^5$). The signal power is also given in the second row for a magnetic field of 12\,T, in place of the present superconducting magnet delivering up to a maximum field of 2\,T. Coupling coefficient $\beta=1$ has been assumed. 
For comparison, a haloscope probing heavier axions ($\nu_c=10\,$GHz corresponds to $41.36\,\,\mu$\,eV) is also considered, having the same pill-box cavity length (0.135\,m). 
}\label{t1}
\end{table}%

The signal power spectrum is a Maxwell-Boltzmann distribution, with $\Delta \nu_a=\nu_a/Q_a$ width, and $Q_a=10^6$ set by axion velocities dispersion in the standard halo model \cite{Turner:1986aa}. 
Note that, though the resonant enhancement is reduced at critical coupling to the receiver chain ($\beta=1$) in a cavity haloscope the optimal coupling maximizes the search speed, and $\beta \sim 2$ or greater is typically chosen.


\subsection{Scan rate}
Given that the axion mass $m_a$ is unknown, the merit of a cavity haloscope design is typically evaluated using the scan rate $df/dt$. This parameter quantifies the maximum speed at which a search can be run at sensitivity $g_{a \gamma \gamma}\propto m_a$ and is given by \cite{Kim_2020}: 

\begin{equation}\label{eq:3}
\frac{df}{dt} \approx \frac{g_{a \gamma \gamma}^4}{\Sigma^2} \frac{\rho_a^2}{m_a^2} \frac{B_{0}^4\, C_{010}^2 V^2}{N_{sys}^2}\left(\frac{\beta }{1+\beta}\right)^2 \frac{Q_L Q_a^2}{Q_L+Q_a} \; \,,
\end{equation}

where $N_{sys}=k_B T_s$ is the system noise \cite{Lamoreaux:2013} with known noise temperature $T_n$. For an amplifier at SQL, $k_B T_s=h\nu$, thus $df/dt \propto \nu^{-4}$ if we fix the cavity length to a value $h$ comparable with the typical length of a commercially available SC solenoid ($\sim 25$\,cm). In fact, $V \propto \nu_c^{-2} $ because in a cylindrical cavity $r=2.405\, c/(2\pi \nu_c)$, $m_a\simeq h\nu_c$, and in addition we drop the quality factor frequency dependence $Q_L\propto \nu^{-2/3}$ related to the anomalous skin effect \cite{pippard:1947}because we consider superconducting cavities. 
Even though we allow for rather extreme values of cavity aspect ratio $h/r\simeq10$, with $r$ cavity radius, the intruder modes density is acceptable in the 5-10\,GHz range, where a few different radius resonators could be envisaged to cover the whole range.  

\noindent

\noindent
\subsection{Photon counting versus linear detection}

The comparison between linear detection and photon counting was thoroughly analyzed in the foundational study by Lamoureaux et al. in Ref.\cite{Lamoreaux:2013}. This analysis is revised and contextualized here, articulated through the lens of experimental parameters, especially focusing on detector efficiency and dark counts. The case where the axion linewidth is smaller than the haloscope linewidth $\Delta \nu_a<\Delta \nu_c$ is considered. In the opposite limit, the analysis holds by substituting the axion linewidth with the haloscope linewidth ($\Delta \nu_a\leftarrow\Delta \nu_c$).

Linear amplifiers are widely utilized for detecting incoherent signals. Consideration is given to an axion signal with power $P_a$, incoherently emitted across a bandwidth $\Delta \nu_a$ at a frequency $\nu_a$, which is predicted by the axion model to typically have a quality factor $Q_a=\nu_a/\Delta \nu_a \sim 10^6$. The noise generated by the linear receiver over a measurement duration $t$ is expressed as:
\begin{equation}
P_{\mathrm{lin}}=h\nu_a (\bar{n}+1) \sqrt{\frac{\Delta \nu_a}{t}}
\end{equation}
where  the mean photon number per mode is given by $\bar{n}=(\exp{(h\nu_a/k_bT)}-1)^{-1}$.

On the one hand, when $k_bT$ significantly exceeds $h \nu_a$, this noise limit simplifies to the Dicke radiometer formula: $P_{\mathrm{RM}}= k_b T \sqrt{\Delta \nu_a/t}$. On the other hand, the optimal noise power is achievable at zero temperature and defined as the standard quantum limit (SQL):
\begin{equation}
P_{\mathrm{SQL}}= h\nu_a \sqrt{\Delta \nu_a/t}
\end{equation}
It is noteworthy that, given the incoherent nature of the expected signal, both I and Q quadratures contribute a vacuum noise of $h\nu_a/2$ leading to noise power of $h\nu_a$ as described in Ref.\cite{Lamoreaux:2013}.
The detection's signal-to-noise ratio (SNR) is therefore capped as follows:

\begin{equation}
\mathrm{SNR}_{\mathrm{SQL}} = \frac{P_a}{h\nu_a} \sqrt{\frac{t}{\Delta \nu_a}}
\end{equation}

In the current setup, we examine a photon detector characterized by an efficiency $\eta$ and a dark count rate $\Gamma$. 
The signal is obtained by accumulating counts over a period of $t$, yielding $S=\eta P_a/h\nu_a  t+\Gamma t$. Given its Poisson distribution, the variance associated with this signal is $\delta S^2=\eta P_a/h\nu_a t+\Gamma t$. 
The SNR therefore reads:
\begin{equation}
\mathrm{SNR}_{\mathrm{PC}}=\frac{\eta P_a t/h\nu_a}{\sqrt{\Gamma_\mathrm{dc} t + \eta P_a t /h\nu_a }}=\frac{\eta P_a}{h\nu_a}\frac{ \sqrt{t} }{\sqrt{\Gamma_\mathrm{dc} + \eta P_a/h\nu_a }}
\end{equation}

In situations where signal contribution $\eta P_a/h \nu_a$ is negligible compared to the darkcount $ \Gamma_\mathrm{dc} $, the noise due to signal shot can be disregarded, leading to the following SNR expression:

\begin{equation}
\mathrm{SNR}_{\mathrm{PC}}\approx\frac{\eta P_a}{ h\nu_a}\sqrt{\frac{ t }{\Gamma_\mathrm{dc} }}
\end{equation}

The darkcount rate can be breakdown in two main contributions $\Gamma_\mathrm{dc}=\Gamma_\mathrm{th}+\Gamma_\mathrm{int}$, where $\Gamma_\mathrm{th}=\eta \Delta \nu_\mathrm{det} n_\mathrm{th} $ is the background thermal fluctuation integrated over the detector bandwidth $\Delta \nu_\mathrm{det}$ and where $\Gamma_\mathrm{int}$ is the intrinsic detector darkcount  due to technical counts due to for instance errors in the qubit readout or out-of-equilibrium excitation of the qubit. In the limit where the darkcount is dominated by thermal background contribution ($\Gamma_\mathrm{int}\ll \eta \Delta \nu_\mathrm{det} n_\mathrm{th}$) the SNR reads

\begin{equation}
\mathrm{SNR}^\mathrm{th}_{\mathrm{PC}}\approx\frac{P_a}{ h\nu_a}\sqrt{\frac{\eta  t }{\Delta \nu_\mathrm{det} n_\mathrm{th} }}
\end{equation}

Note that the SMPD used in this experiment operates in this regime \cite{balembois2023} where the intrinsic darkcount is of the order of $\Gamma_\mathrm{int}\sim 10\ \mathrm{s^{-1}}$ while $\Gamma_\mathrm{th}\sim 75\ \mathrm{s^{-1}}$corresponding to a microwave temperature of $44\ \mathrm{mK}$ where the effective bandwidth of the detector taking into account the Lorentzian linewith is given by $\Delta \nu_\mathrm{det}=\kappa/4=2 \pi/4 \times 0.7\ \mathrm{MHz}$. 

Our goal is to assess the improvement in measurement speed for achieving a specified SNR. 
For operations of a linear receiver at the quantum standard limit, the required measurement time is:
\begin{equation}
t_{\mathrm{SQL}}= \Delta\nu_a \left(\frac{ h \nu_a \mathrm{SNR}  }{P_a}\right)^2
\end{equation}

Conversely, for photon counting the required measurement time is:

\begin{equation}
t_{\mathrm{PC}}= \frac{\Gamma_\mathrm{dc}}{\eta^2} \left(\frac{h \nu_a \mathrm{SNR}}{P_a}\right)^2
\end{equation}

Hence, the comparative gain in measurement speed or scanning rate in this experiment relative to the quantum standard limit is:

\begin{equation}
\mathcal{R}=\frac{t_{\mathrm{SQL}}}{t_{\mathrm{PC}}}=\eta^2\frac{\Delta\nu_a}{\Gamma_\mathrm{dc}}\sim 20.
\end{equation}


The gain in measurement time can be evaluated in the ideal limit where the darkcount is dominated by the background thermal noise and where the detector bandwidth is perfectly matched with the haloscope linewidth ($\Delta\nu_\mathrm{det}\approx\Delta\nu_c$), then the gain is given by:

\begin{equation}
\mathcal{R}^{th}=\frac{t_{\mathrm{SQL}}}{t^\mathrm{th}_{\mathrm{PC}}}=\eta\frac{\Delta\nu_a}{n_\mathrm{th}\Delta\nu_c}.
\end{equation}

Crucially, the enhancement in scan rate achievable through photon counting, as compared to linear detection, is not subject to any fundamental limits. In particular, expected advancements in reducing the dark count rate, enhancing efficiency, and achieving stability (thereby negating the need for the differential method) promise to unlock substantially greater improvements in speed.

\newpage



\begin{figure}[!tbh]
\centering
\includegraphics[width=0.78\textwidth]{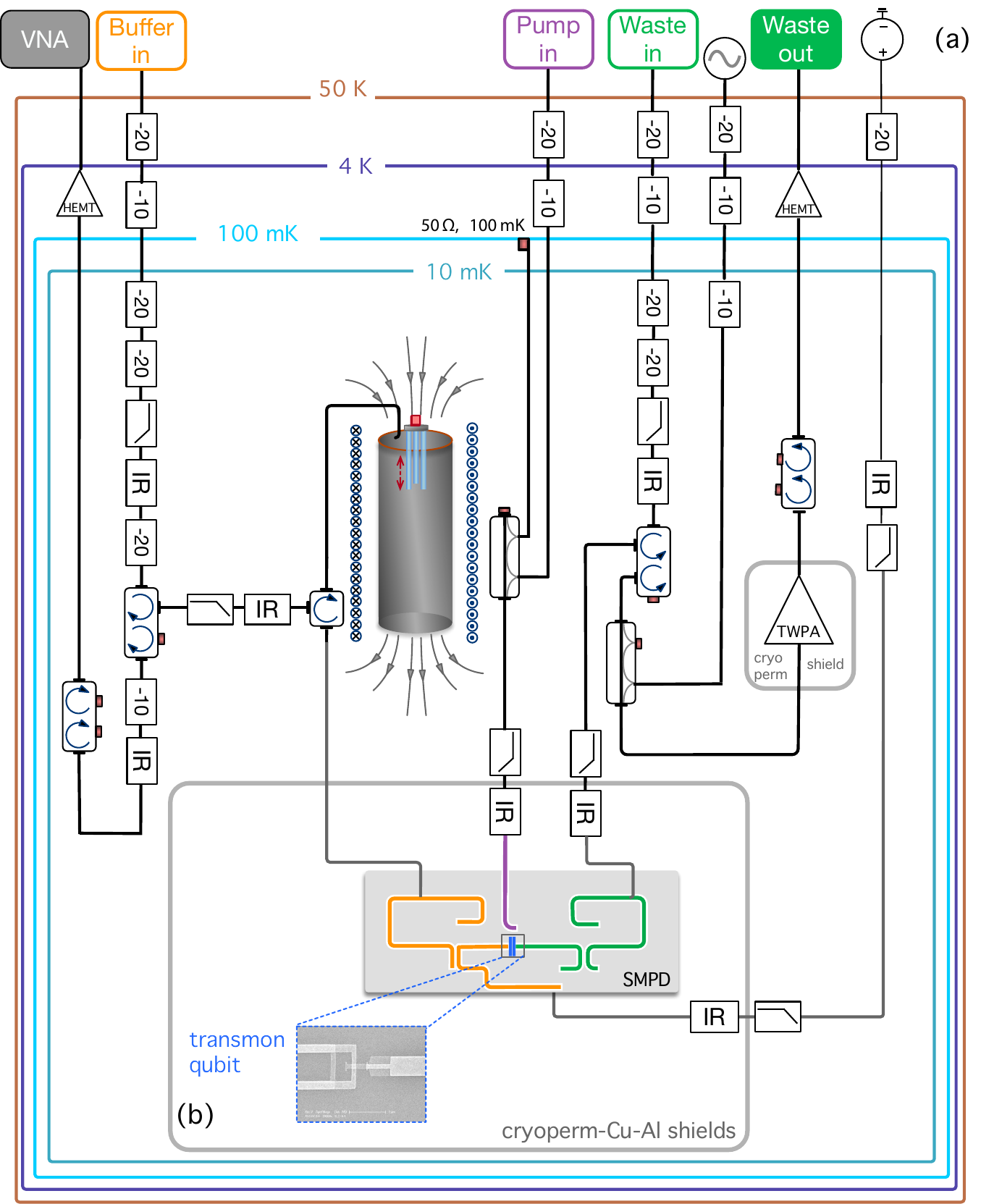}
\caption{
(a) Schematic cabling of the dilution refrigerator. (b) The SMPD is enclosed in a box made with three screens (Al, Cu and cryoperm), while the Traveling Wave Parametric Amplifier (TWPA) is within a separate cylindrical cryoperm shield.}
\label{fig:8}
\end{figure}

\begin{figure}[!tbh]
\centering
\includegraphics[width=.7\textwidth]{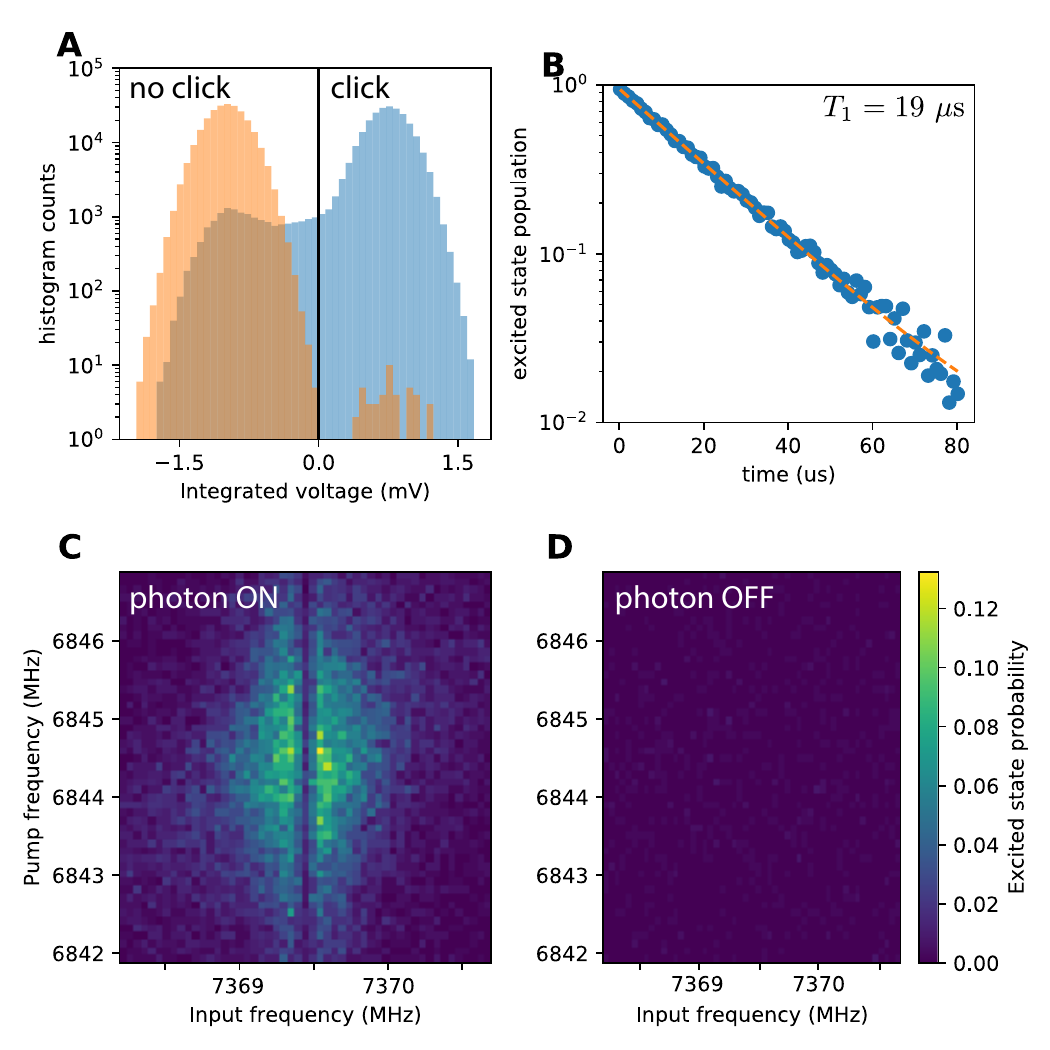}
\caption{\textbf{SMPD Characterization} - \textbf{A:} Readout histogram of the transmon qubit measurement. Readout fidelity for the excited state is $0.93$, while the ground state readout infidelity is less than $5\times10^{-5}$. The thermal equilibrium population of the qubit is $p_\mathrm{th}=2\times10^{-4}$. \textbf{B:} Relaxation measurement of the transmon qubit indicates a $T_1$ value of $19\ \mathrm{\mu s}$. \textbf{C:} Pump frequency tuning involves measuring the qubit state population while driving the pump and a coherent tone on the input resonator. Optimal working points are identified by varying the frequencies of the two drives. \textbf{D:} With the coherent tone turned off, the qubit population remains close to zero, confirming that the correct 4-wave mixing process is at play.}
\label{fig:S1}
\end{figure}

\begin{figure}[!tbh]
 \centering
\includegraphics[width=0.5\textwidth]{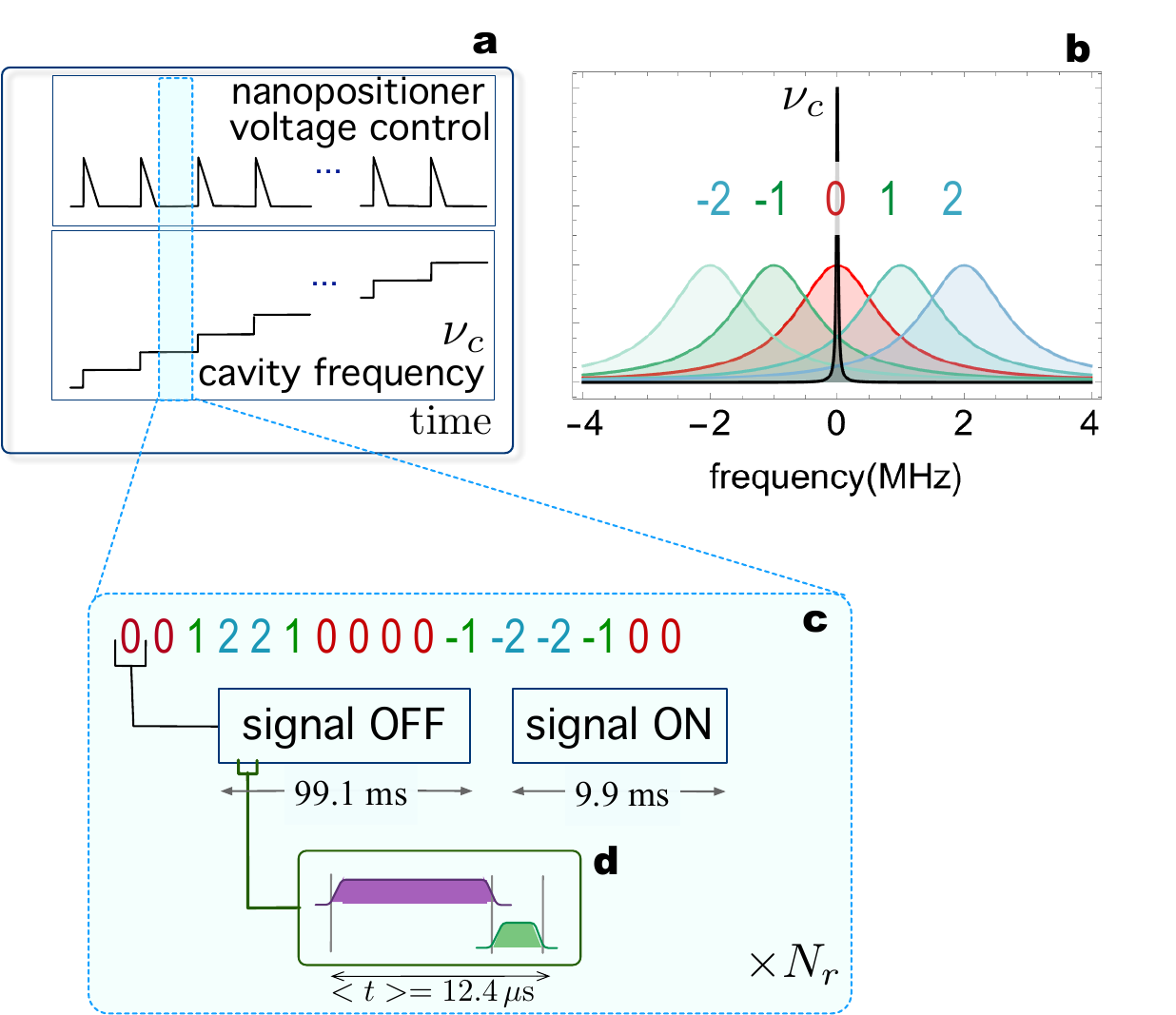}
\caption{
 (a) We tune the cavity by displacing 
 dielectric rods inside the cavity's volume with a linear z-nanopositioner driven by a sawtooth voltage. (b) Values of $\omega_b$ set within the protocol. Labels $-2,-1,1,2$ indicate that clicks are recorded with $\omega_b/2\pi$ differing from the cavity frequency of resonance $\nu_c=\omega_c/2\pi$ by the label numerical value given in MHz. \lq\lq 0" corresponds to the resonance condition $\omega_b=\omega_c$.  To monitor the counter detection efficiency,  white noise is injected during the \lq\lq signal ON'' phase at the buffer input, while with the counts recorded in the \lq\lq signal OFF'' phase the operational dark count can be assessed under different experimental conditions. 
 (c) The buffer frequency $\omega_b/2\pi$ is cycled through the sequence shown for $N_r=36$ times. Before each cycle, a tuning step is devised, in which a weak voltage pulse is applied to the nanopositioner, followed by 5\,s-duration waiting time. 
(d) The detection cycle includes a $10.5\,\mu$s-duration pump pulse (violet), followed by the readout time (green), which is not deterministic. The average duration of this block is 12.4\,$\mu$s.
}
\label{fig:M1}
\end{figure}

\begin{figure}[!tbh]
 \centering
\includegraphics[width=0.8\textwidth]{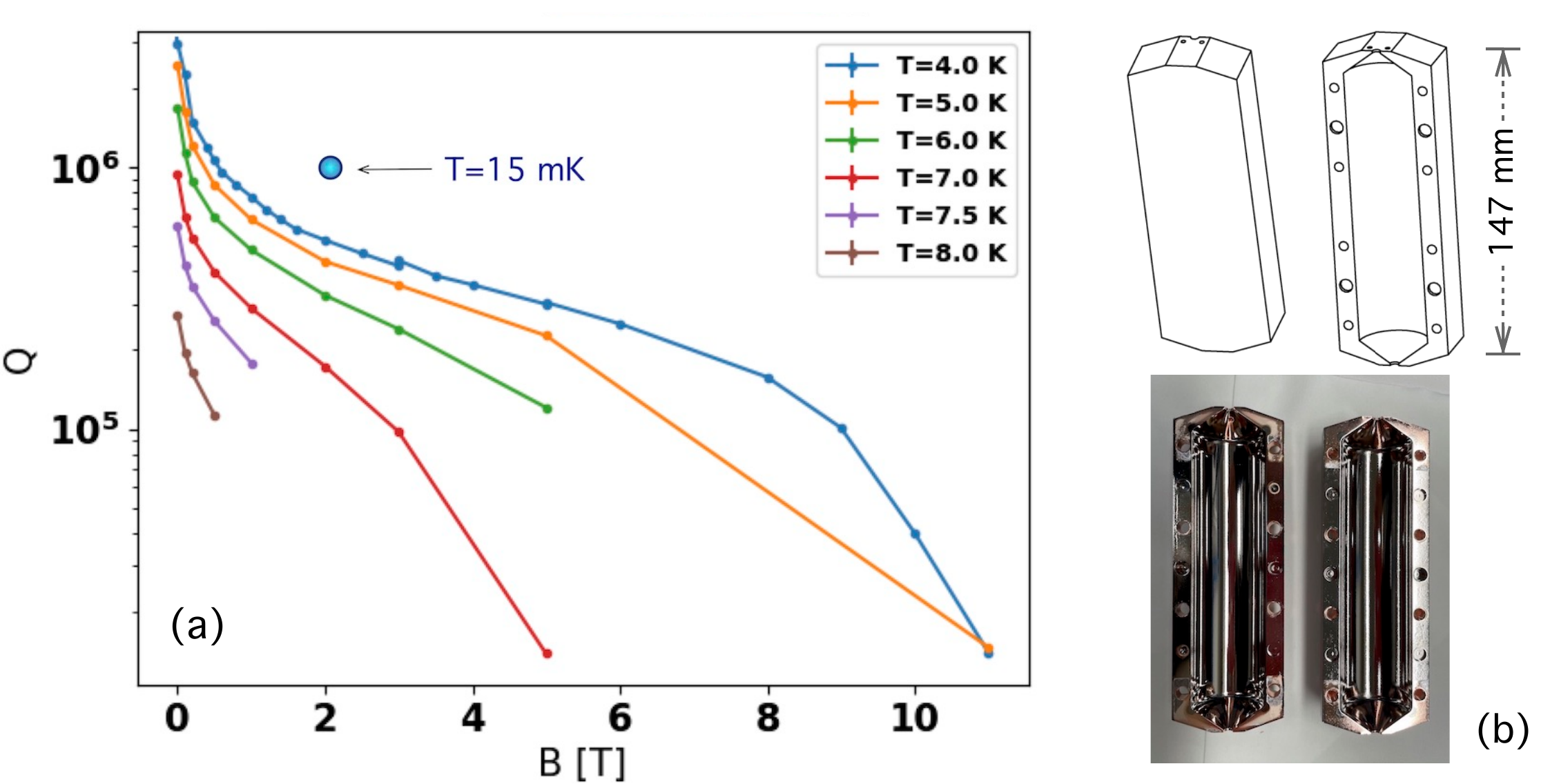}
\caption{
 (a) Microwave cavity quality factor $Q_0$ versus applied magnetic field for a few temperature values, as measured in a Helium flow cryostat. The blue dot is the quality factor measured at 2\,T field in the delfridge at about 15\,mK base temperature. (b) Cavity geometry, realized in two hollowed out copper bodies. Only the central cylindrical part is covered with a superconducting NbTi film. The cavity endcaps are shaped as cones. 
}
\label{fig:M2}
\end{figure}

\begin{figure}[!tbh]
 \centering
\includegraphics[width=0.5\textwidth]{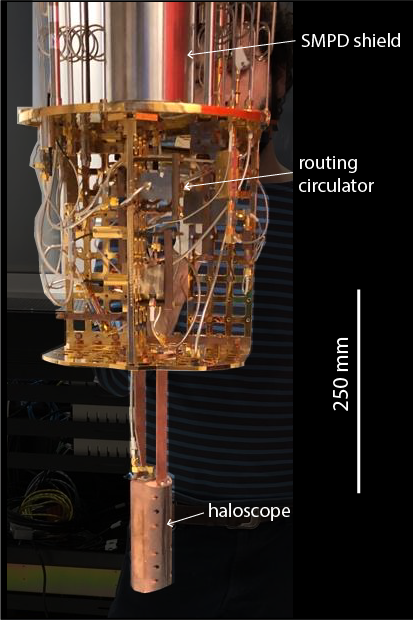}
\caption{
 Photo of the cryostat at room temperature. The haloscope is positioned at the bottom of the cryostat, which operates at millikelvin temperatures, with a superconducting magnet to be mounted around it at 4K. The magnetic shield for the SMPD is situated on the mixing chamber plate, approximately 50 cm from the magnet's center. A circulator directs the signal from the haloscope to the SMPD.
}
\label{fig:M3}
\end{figure}

\begin{figure}[!tbh]
\centering
\includegraphics[width=.7\textwidth]{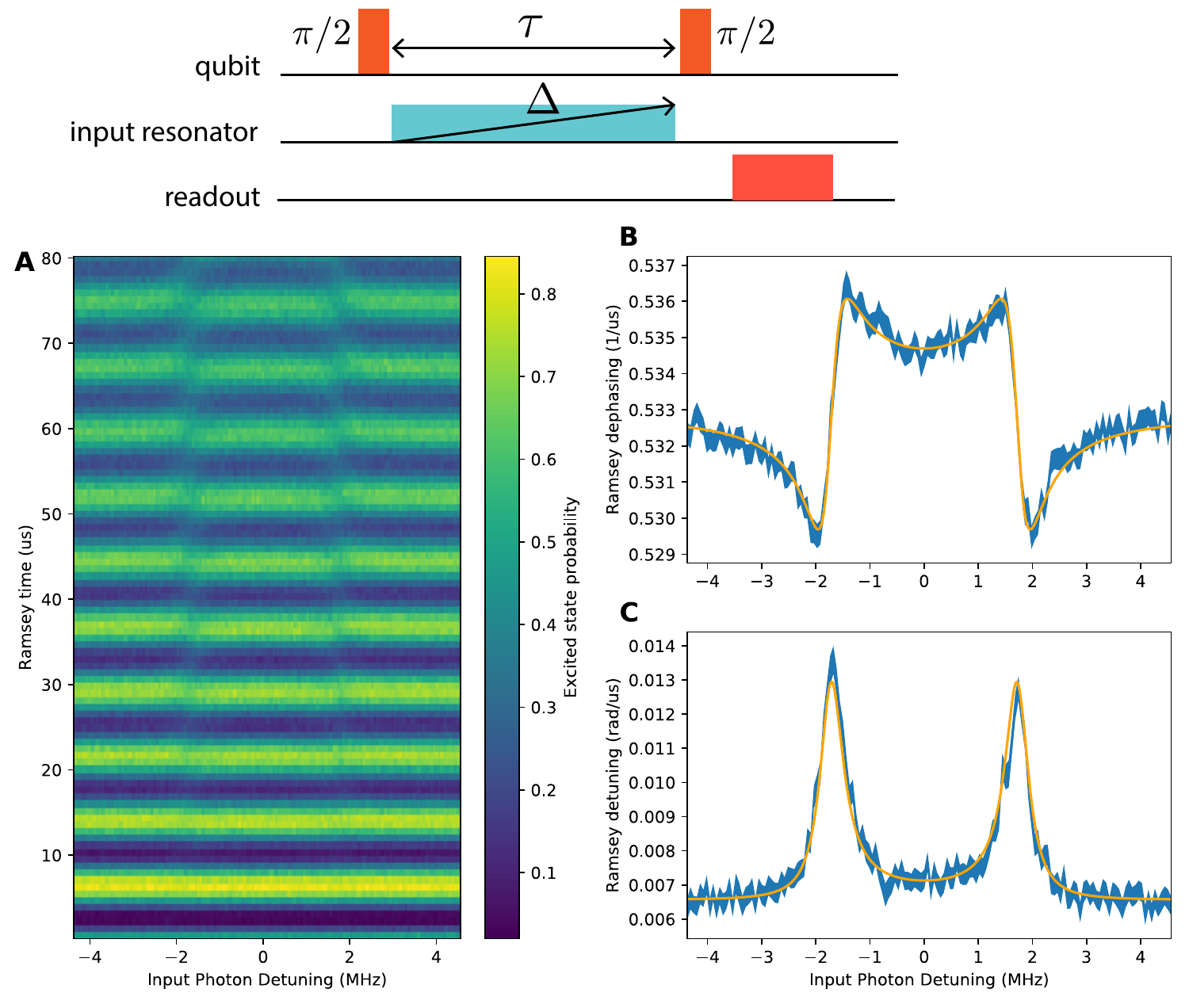}
\caption{\textbf{Coherent Tone Calibration by Ramsey Measurement Under Illumination.} The pulse sequence consists of two $\pi/2$ pulses separated by a Ramsey interval $\tau$. A coherent drive populates the input resonator during the waiting time. Following the sequence, the qubit is read out. \textbf{A:} Qubit population as a function of the waiting time and the frequency of the coherent drive. Ramsey oscillations are observed in the qubit population. When the drive resonates with the input resonator, both a frequency shift and an accelerated decay of the Ramsey oscillations occur. \textbf{B and C:} Simultaneous determination of the frequency shift and decay rate as a function of drive detuning (blue line) by fitting individual Ramsey traces. The width of the line represents the uncertainty in the fit results. The orange line corresponds to the theoretical model described in the main text, enabling the determination of the resonator decay rate ($\kappa/2\pi=0.523\pm 0.014\ \mathrm{MHz}$), the dispersive shift ($\chi/2\pi=3.461\pm 0.015\ \mathrm{MHz}$), and the coherent drive rate ($\epsilon/2\pi=40.9 \pm 0.5\ \mathrm{kHz}$).}
\label{fig:S2}
\end{figure}

\begin{figure}[!tbh]
\centering
\includegraphics[width=.7\textwidth]{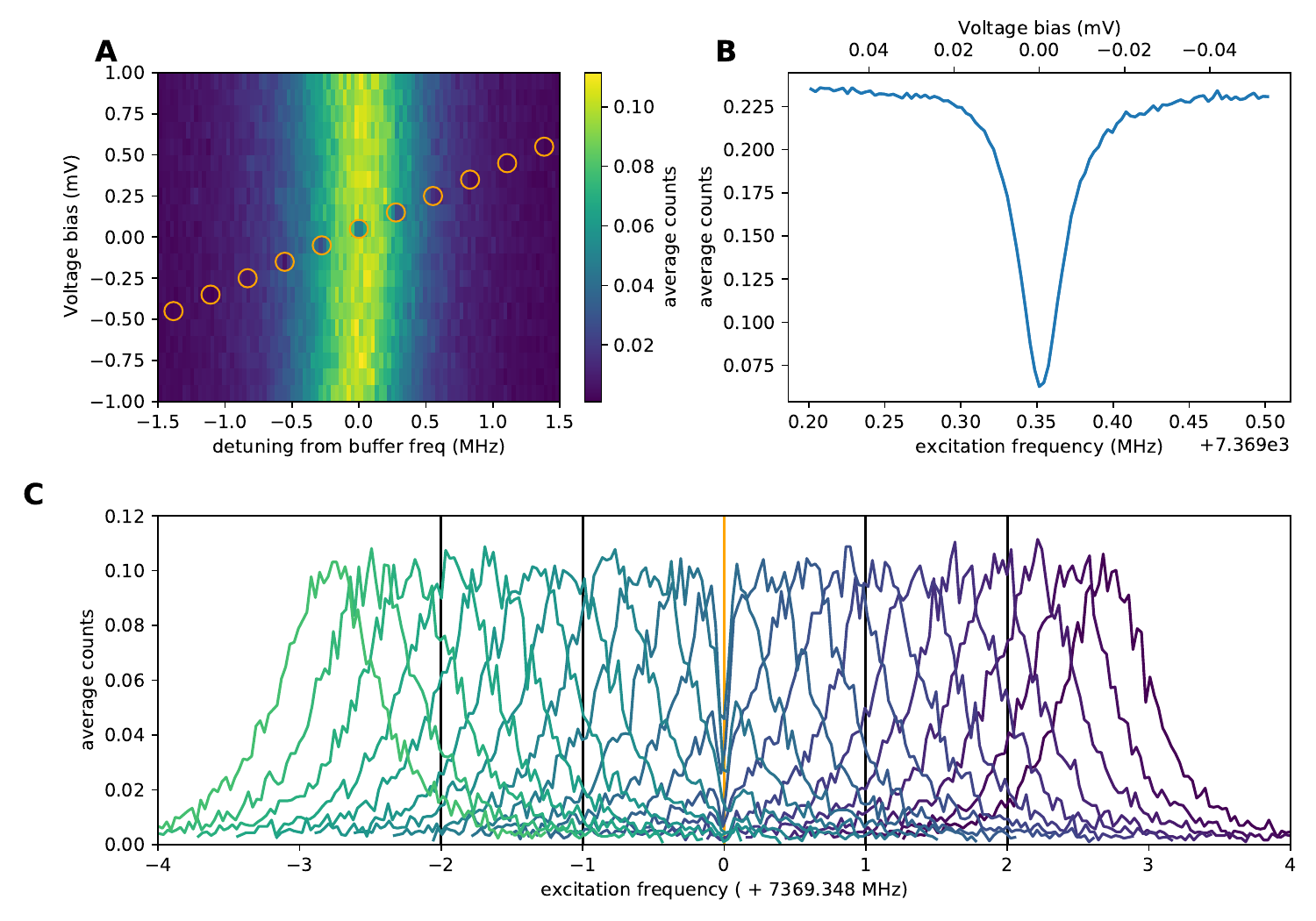}
\caption{\textbf{Frequency Calibration Routine.} \textbf{A:} The click count is measured as a function of the SQUID bias voltage and the detuning between the coherent drive and the expected input resonator frequency at each SQUID bias voltage. Pump frequency adjustments accompany each SQUID bias voltage change. \textbf{B:} The same data presented as a function of the absolute frequency of the coherent drive. Note the sharp decrease in the count number (circled in orange in A), indicating the absorption by the haloscope cavity. \textbf{C:} Both the SMPD frequency and the coherent drive frequency are scanned across the haloscope resonance. The absorption dip observed in the reflection coefficient enables the determination of the haloscope's frequency and linewidth.}
\label{fig:S3}
\end{figure}

\begin{figure}[!tbh]
\centering
\includegraphics[width=1\textwidth]{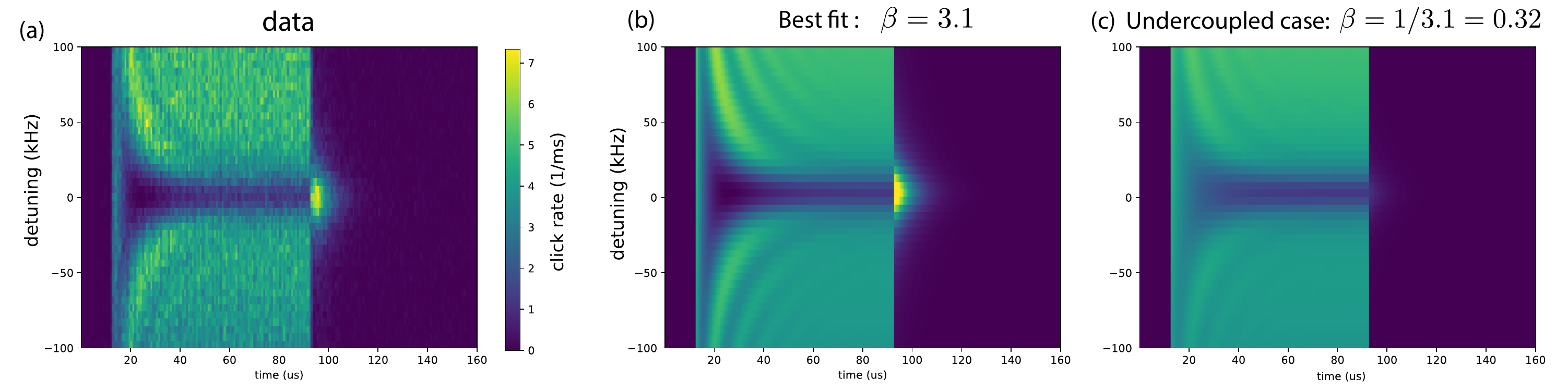}
\caption{\textbf{Time-Domain Response of the Haloscope Cavity Measured with the SMPD.} (a). The click rate of the SMPD is measured in response to a square pulse applied to the haloscope. This measurement is conducted as a function of time and pulse frequency. The SMPD's time resolution is enhanced to $1\ \mathrm{\mu s}$ by reducing the duration of the detection windows. The detection times of the SMPD are sampled to cover the entire time window. (b). Theoretical prediction fitted with $\beta=3.15$ for the haloscope.  All transient features are quantitavely captured by the model lifting the ambiguity toward the overcoupled regime.(c) Theoretical prediction for to the same loaded quality factor $Q_L$ but in the undercoupled regime with $\beta=1/3.15=0.32<1$. The transient features at the loading and unloading of the haloscope are not reproduced despite that the steady-state absorption is identical to the over-coupled case as expected.}
\label{fig:S4}
\end{figure}

\end{document}